\documentclass[prb,twocolumn,amsmath,amssymb]{revtex4}

\usepackage{graphicx}
\usepackage{dcolumn}
\usepackage{bm}
\usepackage{color}
\usepackage{pifont}

\begin{document}
\title{Monitoring surface resonances on Co$_2$MnSi(100) by spin-resolved photoelectron spectroscopy}
 
\author{J. Braun$^1$, M. Jourdan$^2$, A. Kronenberg$^2$, S. Chadov$^3$, B. Balke$^4$, M. Kolbe$^2$, A. Gloskovskii$^5$,
H. J. Elmers$^2$, G. Sch\"onhense$^2$, C. Felser$^{3,4}$,
M. Kl\"aui$^2$, H. Ebert$^1$ and J. Min\'ar$^{1,6}$}

\affiliation{$^1$Department Chemie, Ludwig-Maximilians-Universit\"at M\"unchen, 81377 M\"unchen, Germany \\
$^2$Institut f\"ur Physik, Johannes-Gutenberg-Universit\"at Mainz, Staudingerweg 7, 55128 Mainz, Germany \\
$^3$Max-Planck-Institut f\"ur Chemische Physik fester Stoffe, 01187 Dresden, Germany \\
$^4$Institut f\"ur Anorganische und Analytische Chemie, Johannes-Gutenberg-Universit\"at Mainz, Staudingerweg 9, 55128 Mainz, Germany \\
$^5$Deutsches-Elektronen-Synchrotron DESY, 22603 Hamburg, Germany \\
$^6$New Technologies - Research Center, University of West Bohemia, Univerzitni 8, 306 14 Pilsen, Czech Republic}

\begin{abstract}
The magnitude of the spin polarization at the Fermi level of ferromagnetic materials at room temperature is a key property
for spintronics. Investigating the Heusler compound Co$_2$MnSi a value of 93$\%$ for the spin polarization has been observed
at room temperature, where the high spin polarization is related to a stable surface resonance in the majority band extending
deep into the bulk. In particular, we identified in our spectroscopical analysis that this surface resonance is embedded in
the bulk continuum with a strong coupling to the majority bulk states. The resonance behaves very bulk-like, as it extends
over the first six atomic layers of the corresponding (001)-surface. Our study includes experimental investigations, where
the bulk electronic structure as well as surface-related features have been investigated using spin-resolved photoelectron
spectroscopy (SR-UPS) and for a larger probing depth spin-integrated high energy x-ray photoemission spectroscopy (HAXPES).
The results are interpreted in comparison with first-principles band structure and photoemission calculations which consider
all relativistic, surface and high-energy effects properly.
\end{abstract}


\maketitle

\section{Introduction}
For many spintronics applications it is not the bulk, but the surface or interface electronic structure at the Fermi energy given
by the difference of the normalized total number of spin up and spin down electrons of the involved materials, which is relevant
for applications. However, theoretical predictions for this sample region are much more demanding compared with calculations of
bulk properties. A key property for spintronics is the spin polarization at the Fermi level. Concerning surface and interface
states it is necessary to distinguish between materials with a finite total spin polarization at the Fermi energy and materials
with zero total but momentum dependent spin polarization, which is nonzero for specific k values. Examples for the second class
of materials are topological insulators like Bi$_2$Se$_3$ \cite{Hsi09}, but also simple materials like Bismuth \cite{Tak11} and
Tungsten \cite{Miy12}. However, the first class of materials, i.e. with a nonzero total spin polarization, can be realized by
ferro- or ferrimagnetic materials only. Accordingly, in the following the term spin polarization is always used for the total
spin polarization at the Fermi energy. By surface sensitive spin and angular-resolved photoemission spectroscopy (SR-ARPES) values
of the spin polarization close to 100$\%$ at room temperature were observed for metastable CrO$_2$ \cite{Kam87} and for Fe$_3$O$_4$
\cite{Ded02}, but no corresponding state-of-the-art photoemission calculation and no discussion of possible surface states exists
up to now. Furthermore, these materials did not allow for large spin transport effects and are not compatible with other materials
relevant for applications. For these reasons intermetallic Heusler compounds \cite{Gra11} with their predicted half-metallic
properties \cite{Gro83} moved into the focus of interest \cite{Plo99,Kol05,Don05,Cor06,Yab06,Min09}. In addition to being interesting
for applications, intermetallic Heusler materials represent a test for modern electronic structure calculations for materials with
electronic correlations of moderate strength \cite{Kat08,Cha09,Wus12,Bra13}. In fact, by means of various band structure methods
many Heusler compounds have been predicted to be 100$\%$ spin-polarized in the bulk. However, the first direct observation of a
huge surface spin polarization in any Heusler compound by photoemission spectroscopy was possible only very recently \cite{Jou14}.
In our preliminary work we identified a 93$\%$ polarized surface resonance investigating the Heusler compound Co$_2$MnSi at room
temperature. This represents the first example for such a resonance in any ferromagnetic metal. Within this work we investigate
in more detail the occupied as well as the unoccupied electronic structure of Co$_2$MnSi. Furthermore, we compare our
spectroscopical analysis to corresponding experimental data, with special emphasis on surface-related features of the electronic
structure.

\section{Theoretical and experimental details}
\subsection{Theoretical and computational details}
Experimentally, the interesting valence band region around the Fermi energy is accessible by means of ultraviolet photoemission
spectroscopy (PES) \cite{pes} and inverse photoemission spectroscopy (IPE) \cite{ipe}. From the theoretical point of view the
most successful theoretical approach to deal with photoemission is the so-called one-step model as originally proposed by Pendry
and co-workers \cite{Pen74,Pen76,HPT80}. A review on the various developments and refinements \cite{dev} of the approach can be
found in Ref.~\onlinecite{Bra96}. As the most recent development our spectroscopic analysis is based on the fully relativistic one-step
model, in its spin-density matrix formulation. This approach allows describing properly the complete spin-polarization vector
of the photo current. The corresponding spin-density matrix of the photocurrent is defined by the following equation \cite{Halil93}:
\begin{equation}
{\overline\rho}^{\rm PES}_{ss'}({\bf k}_{||},\epsilon_f) =  \langle s, \epsilon_f, {\bf k}_{\|} | G_{2}^+ \Delta G_{1}^+
\Delta^\dagger G^-_{2} | \epsilon_f, {\bf k}_{||}, s' \rangle~.
\label{eq:spind1}
\end{equation}
It follows then for the spin-density matrix $\rho$:
\begin{equation}
\rho^{\rm PES}_{ss'}({\bf k}_{||},\epsilon_f)~=~\frac{1}{2i}~\left({\overline\rho}^{\rm PES}_{ss'}({\bf k}_{||},\epsilon_f)~-~
{\overline\rho}^{*~\rm PES}_{s's}({\bf k}_{||},\epsilon_f) \right)~.
\label{eq:spind2}
\end{equation}
The intensity of the photocurrent results in:
\begin{equation}
I^{\rm PES}({\bf k}_{||},\epsilon_f)~=~Sp \left(~\rho^{\rm PES}_{ss'}({\bf k}_{||},\epsilon_f)~\right)~,
\label{eq:spind3}
\end{equation}
and the corresponding spin-polarization vector is given by:
\begin{equation}
{\bf P}~=~\frac{1}{I}~Sp~\left(~ \mbox{\boldmath $\sigma$} \cdot \rho~\right)~,
\label{eq:spind4}
\end{equation}
where $\mbox{\boldmath $\sigma$}$ denotes the vector of the three Pauli spin matrices. Finally, the spin-projected photocurrent
is obtained from the following equation:
\begin{equation}
I^{\pm~\rm PES}_{{\bf n}}~=~\frac{1}{2}~\left(~1~\pm~{\bf n} \cdot {\bf P}~\right)~I^{\rm PES}.
\label{eq:spind5}
\end{equation}
The spin polarization is calculated with respect to the vector {\bf n}. This, for example, allows the complete calculation of
all three components of the spin-polarization vector for each pair of (k$_x$,k$_y$) values which define the coordinate system for
momentum images. Within this formalism I$^{PES}$ denotes the elastic part of the photocurrent. Vertex renormalizations are neglected.
This excludes inelastic energy losses and corresponding quantum-mechanical interference terms \cite{Pen76,Bor85,Car73}. Furthermore,
the interaction of the outgoing photoelectron with the rest system is not accounted for, which means that the so-called sudden
approximation has been applied. This approximation is expected to be justified for not too small photon energies. The initial and
final states are constructed within spin-polarized low-energy electron diffraction (SPLEED) theory where the final state is
represented by a so-called time-reversed SPLEED state \cite{Bra96,Bra01}. Many-body effects are included phenomenologically in the
final-state calculation, using a parameterized, weakly energy-dependent and complex inner potential as in Ref.~\onlinecite{Pen74}. This
generalized inner potential accounts for inelastic corrections to the elastic photocurrent \cite{Bor85} as well as the actual
(real) inner potential, which serves as a reference energy inside the solid with respect to the vacuum level \cite{Hil95}. Due to
the finite imaginary part, the inelastic mean free path (IMFP) is accounted for and thus the amplitude of the high-energy photoelectron
state can be neglected beyond a certain distance from the surface. 

The self-consistent electronic structure calculations were performed within the ab-initio framework of spin-density functional 
theory, in a fully relativistic mode by solving the corresponding Dirac equation. For the exchange and correlation potential the
Perdew-Burke-Enzerhof parametrization was used \cite{Per96}. To account for electronic correlations beyond the local spin-density
approximation (LSDA) we employed a self-consistent combination of the LSDA and the dynamical mean field theory. This computational
LSDA+DMFT scheme, self-consistent in both the self-energy calculation and the charge-density calculation is implemented within the
relativistic SPR-KKR formalism \cite{Min11,MCP+05,Ebe12,Ebe121}. The effective DMFT impurity problem was solved through the spin
polarized T-matrix fluctuation-exchange (SPTF) solver \cite{PKL05}, working on a Matsubara energy grid corresponding to a temperature
of 400 K. The SPTF solver is accurate for moderately correlated systems as it was shown in several successful applications for
various materials \cite{Min11}. Simultaneous convergence of the electronic charge-density and the self-energy had been achieved by
use of 4096 Matsubara frequencies. The double counting was corrected using the fully localised limit (FLL) scheme (for details
concerning DC corrections within KKR calculations see \cite{Min11}). The FFL was succesfully applied to Co$_2$MnSi recently within
LSDA+U calculations \cite{MFR+12},where we used the fully rotationally invariant U-matrix  with U-parameters U$_{\rm Mn}$=3.0 eV
and U$_{\rm Co}$=1.5 eV for Mn and Co. The exchange parameter $J$ was chosen to 0.9 eV for both Mn and Co.
 
Additionally, for the photoemission calculations, we account for the surface barrier by use of a Rundgren-Malmstr\"om-type surface
potential \cite{Mal80}, which can be easily included into the formalism as an additional layer. This procedure allows for the
correct description of its asymptotic behavior. As this surface barrier represents a z-dependent potential, a surface contribution
as part of the total photocurrent results, which accounts explicitly for the energetics and dispersion of all surface features.
Furthermore, the relative intensities of surface-related spectral distributions are quantitatively accounted for by calculating the
corresponding matrix elements in the surface region. This procedure is described in detail, for example, in Refs.~\onlinecite{Nub11,Ben11}.
Also, energy and momentum conservation are naturally included in the formalism \cite{Hop80,Bra96}. To take care of impurity scattering,
a small constant imaginary value of V$_{i1}$ = 0.05 eV was used for the initial state, this way describing the finite life-time of
the initial state. Life-time effects in the final state are accounted for by the imaginary part of the inner potential (see above).
A constant imaginary value of V$_{i2}$ = 1.5 eV has been chosen again in a phenomenological way for excitation energies in the
ARPES regime. According to the experimental setup the spectroscopic calculations were performed for linearly p-polarized light.

\subsection{Experimental details}
The high reactivity of Heusler materials makes photoemission spectroscopy (PES), the most powerful method for investigations of the
electronic band structure of solids, challenging. Sample degradation is a major problem and often results in missing Fermi edges in
the photoemission spectra and unexpectedly small experimentally obtained spin polarizations, such as 12$\%$ for the predicted half-metal
Co$_2$MnSi \cite{Wan05}. Sputter cleaning of the samples prior to PES yields slightly increased spin polarization values, for instance
20$\%$ for Co$_2$Cr$_{0.6}$Fe$_{0.4}$Al \cite{Wus09}. However, in these and many other cases neither the observed spin polarization
nor the overall energy dependence of the total intensity resembles the results of band structure calculations \cite{Wur06}. Generally
three issues result in poor agreement of theory with experiment: Imperfect surface preparation often leads to disorder, aging of the
sample results in surface oxidation, and the discrepancies with calculated spectra may arise from shortcomings of the calculations
themselves, e.g. the neglect of surface states. However, calculations of surface states of non-stoichiometric Co$_2$MnSi(100) thin
films are available \cite{Wus12}, which were used to explain the low spin polarization of about 20$\%$ measured by UPS (h$\nu$=5.9 eV)
on ex-situ prepared and in-situ sputter cleaned Co$_2$Mn$_{1.19}$Si$_{0.88}$ thin films. Fetzer et al. \cite{Fet13} published
experiments of ex-situ prepared stoichiometric Co$_2$MnSi(100) capped by 20 monolayers of MgO through which they obtained spin
resolved UPS data (h$\nu$=5.9 eV). They identified no interface states, which they attributed to defects at the Co$_2$MnSi/MgO
interface. A spin polarization of about 40$\%$ was obtained. Another way to avoid spurious photoemission results due to degraded
sample surfaces is the use of less surface-sensitive hard x-ray photoemission spectroscopy (HAXPES). The identification of several
experimentally obtained intensity features with the density of states (DOS) was possible for materials like NiTi$_{0.9}$Sc$_{0.1}$Sn,
NiMnSb or Co$_{x}$Mn$_{y}$Ge$_{z}$ (x : z = 2 : 0.38) thin Heusler films \cite{Qua11,Qua13a}. However, due to the low intensities
of HAXPES experiments no spin resolved results are available up to now. We demonstrated that by in-situ UPS with highly efficient
spin-filtering \cite{Kol11} on epitaxial Heusler thin films the problem of surface degradation can be solved and a spin polarization
of 55$\%$ investigating Co$_2$MnGa could be obtained \cite{Kol12}. Very recently, we measured a record value of 93$\%$ at room
temperature investigating Co$_2$MnSi by this method \cite{Jou14}.

\section{Discussion}
\begin{figure}[tp]
\includegraphics[width=0.27\textwidth,clip]{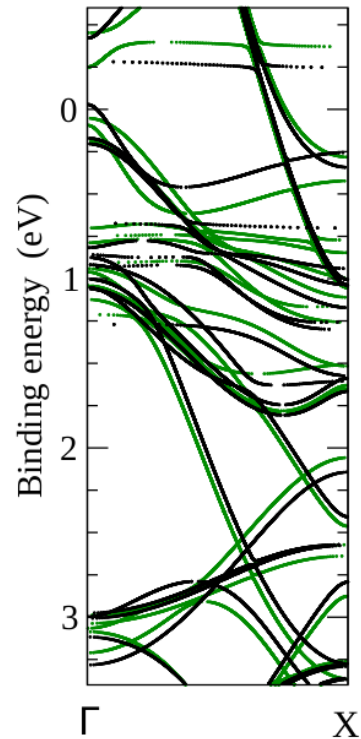}
\caption {Relativistic bulk band structure calculated along the $\Delta$-direction for a magnetization direction perpendicular
to the (001)-surface (black color) and for an in-plane magnetization direction along $\overline \Gamma$-$\overline M$ (green,
light color).}
\label{figure1}
\end{figure}
In Fig.~1 we present the fully-relativistic band structure calculated for the $\Delta$-direction of the bulk Brillouin zone.
The Co$_2$MnSi-films are magnetized in-plane along the $\overline \Gamma$-$\overline M$ line of the surface Brillouin zone.
However, we have calculated additionally the bulk-band dispersions for a magnetization direction perpendicular to the surface
to demonstrate that the dispersion strongly depends on the magnetization direction. This behavior results from the interplay
between spin-orbit-coupling (SOC) and magnetic exchange interaction and depends on the band symmetry \cite{DKS+94}. Here, the
magnetization direction perpendicular to the surface is indicated by black color and the in-plane direction as given from the
experimental conditions is marked by green (light) color. As the underlying symmetry is different for these two cases the
spin-orbit coupling causes pronounced changes in the dispersion and energetic position of nearly all bands, if the magnetization
is switched from in-plane to out-of-plane direction. The corresponding energy-band splittings and hybridization gaps typically
range from some meV up to about 100~meV. In contrast the calculated magnetocrystalline anisotropy energy (MAE), which is
defined as the energy difference between $\Delta$E$_{MAE}$ = E$_{M||[001]}$-E$_{M||[110]}$ of bulk Co$_2$MnSi results to
$\Delta$E$_{MAE}$ = 0.4 $\mu$eV. This should be expected for bulk Co$_2$MnSi due its cubic structure. The reason for the
different energy scales appearing here is found in the fact that the SOC-split energy-band regions only contribute to the
MAE if they are located in the vicinity of the Fermi level, where they induce significant deformations of the Fermi surface
As a consequense they do not contribute to the MAE if they appear at finite binding energies. Note that similar effects can be
observed for a magnetization which is directed parallel to the (100) and (001) crystallographic axis. The energetic difference
vanishes for the bulk system because these two axis are equivalent. In this case the electronic band structure is invariant
under rotations of the quantization axis together with the magnetization direction \cite{Stoe99}. As discussed before, this
means that all these modifications cancel each other after integration over all occupied states and do not contribute (as
in the cases of (100) and (001) directions) or only create very small contributions (in the present case of (110) and (001)
directions) to the magnetocrystalline anisotropy.   

One should expect that also surface-related features of the electronic structure will be significantly influenced if the
magnetization direction changes. This is indeed the case and will be discussed later in context with the nearly 100$\%$ spin
polarization which was found for in-plane magnetized samples. 
\begin{figure}[bp]
\includegraphics[width=0.20\textwidth,clip]{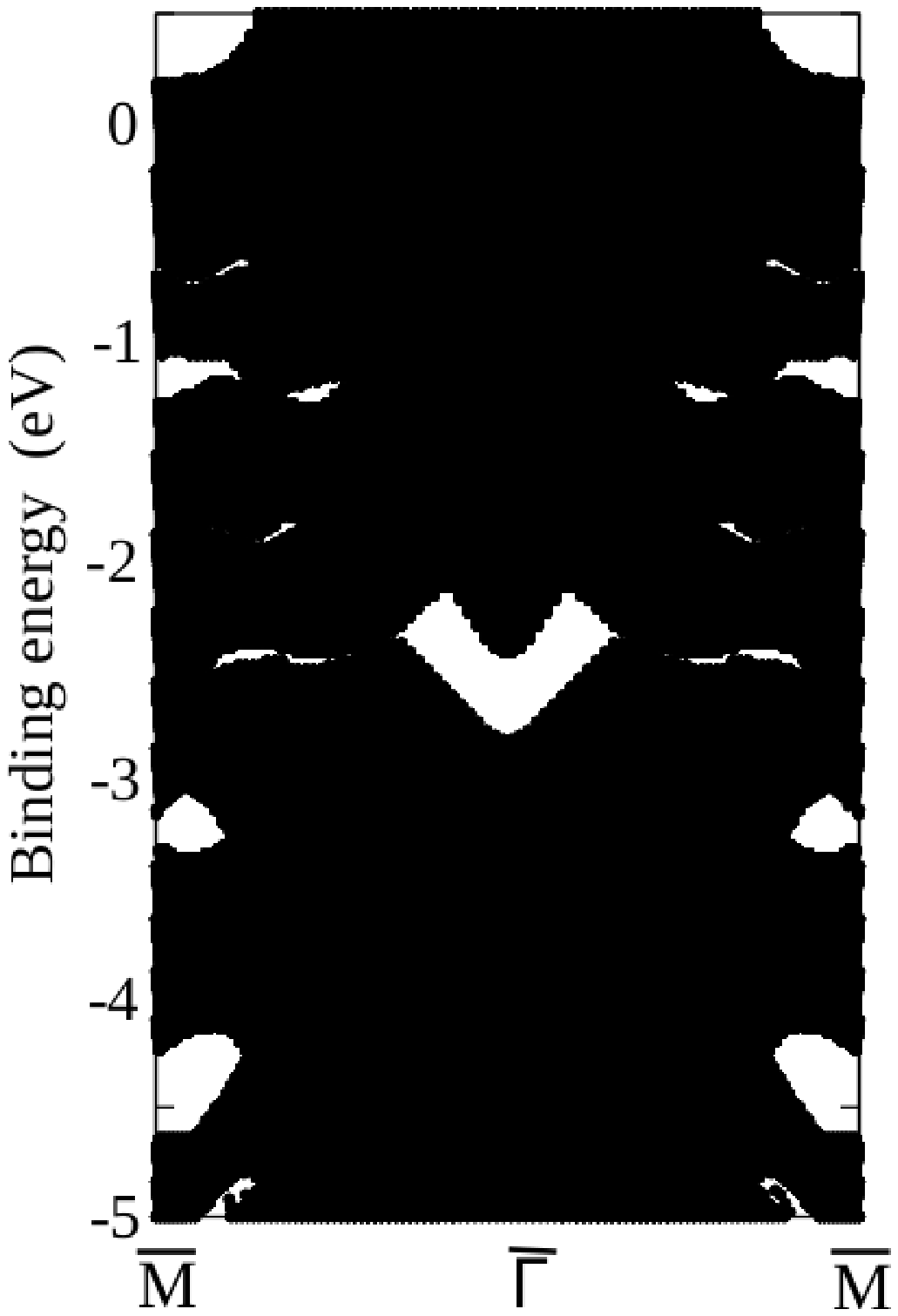}
\includegraphics[width=0.195\textwidth,clip]{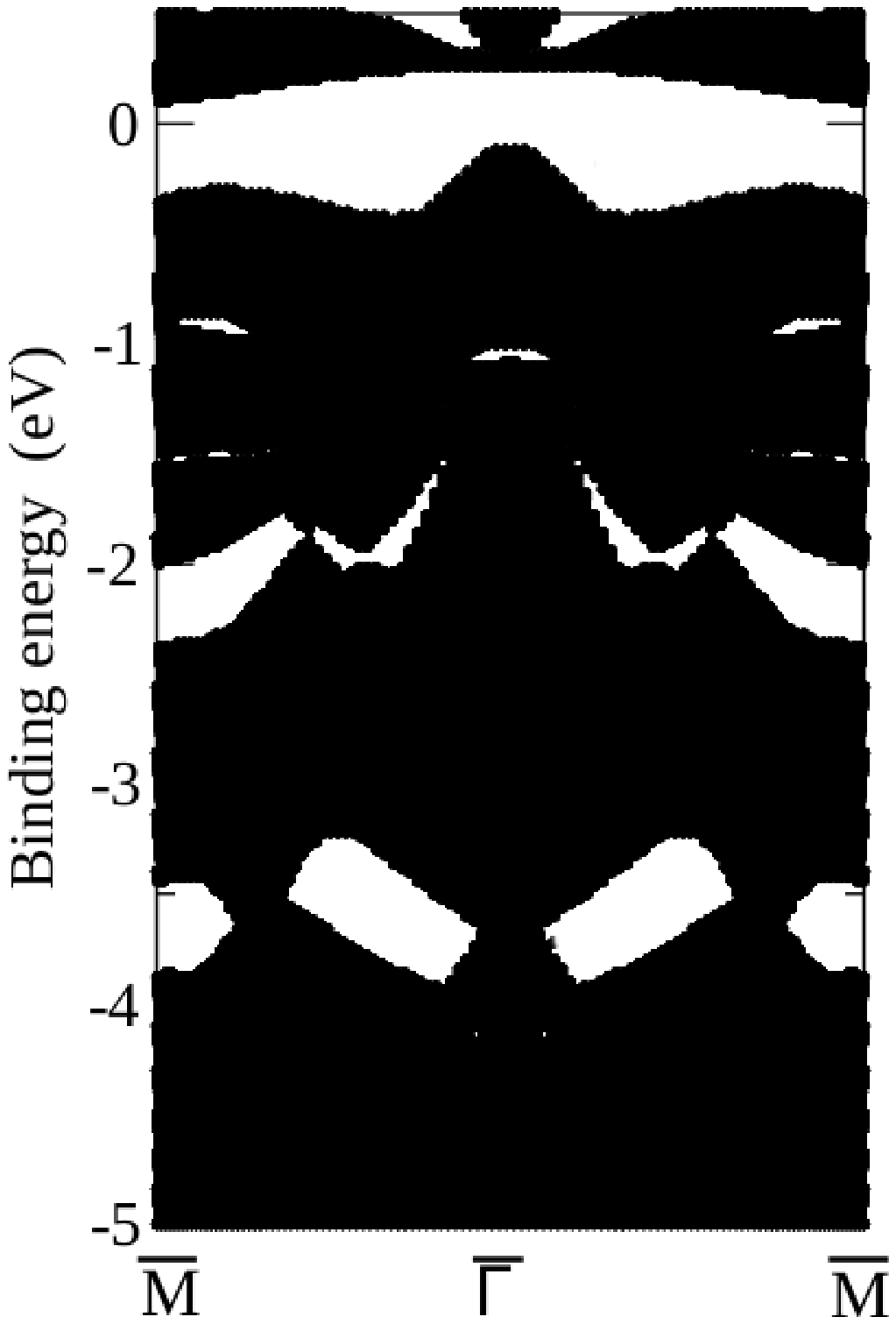}
\caption {Projected relativistic bulk band structures for majority spin (left panel) and minority spin character (right panel)
along the $\overline \Gamma$-$\overline M$ direction. Dark colored regions represent  the projection of bulk states.}
\label{figure2}
\end{figure}
\begin{figure}[tp]
\includegraphics[width=0.23\textwidth,clip]{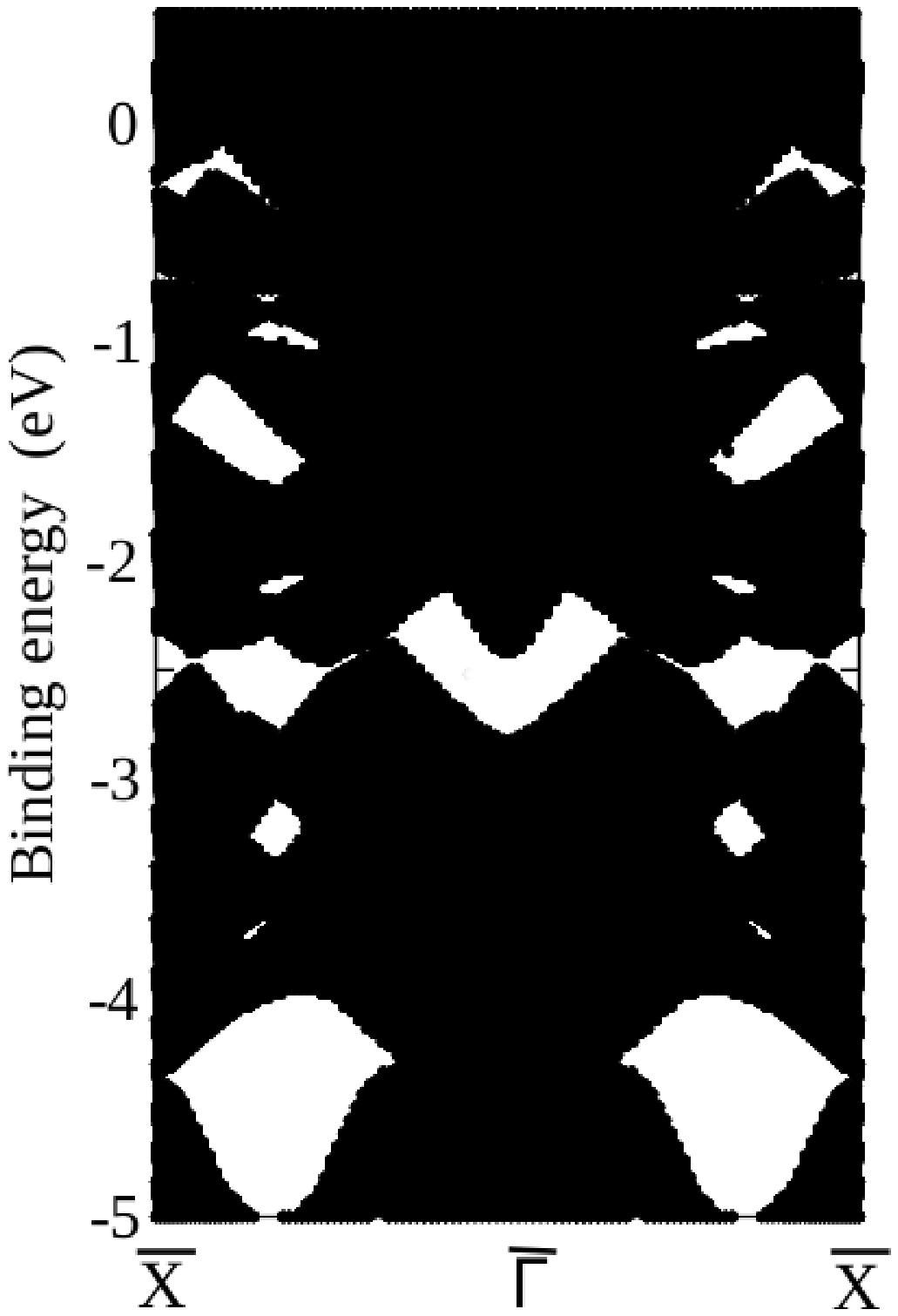}
\includegraphics[width=0.23\textwidth,clip]{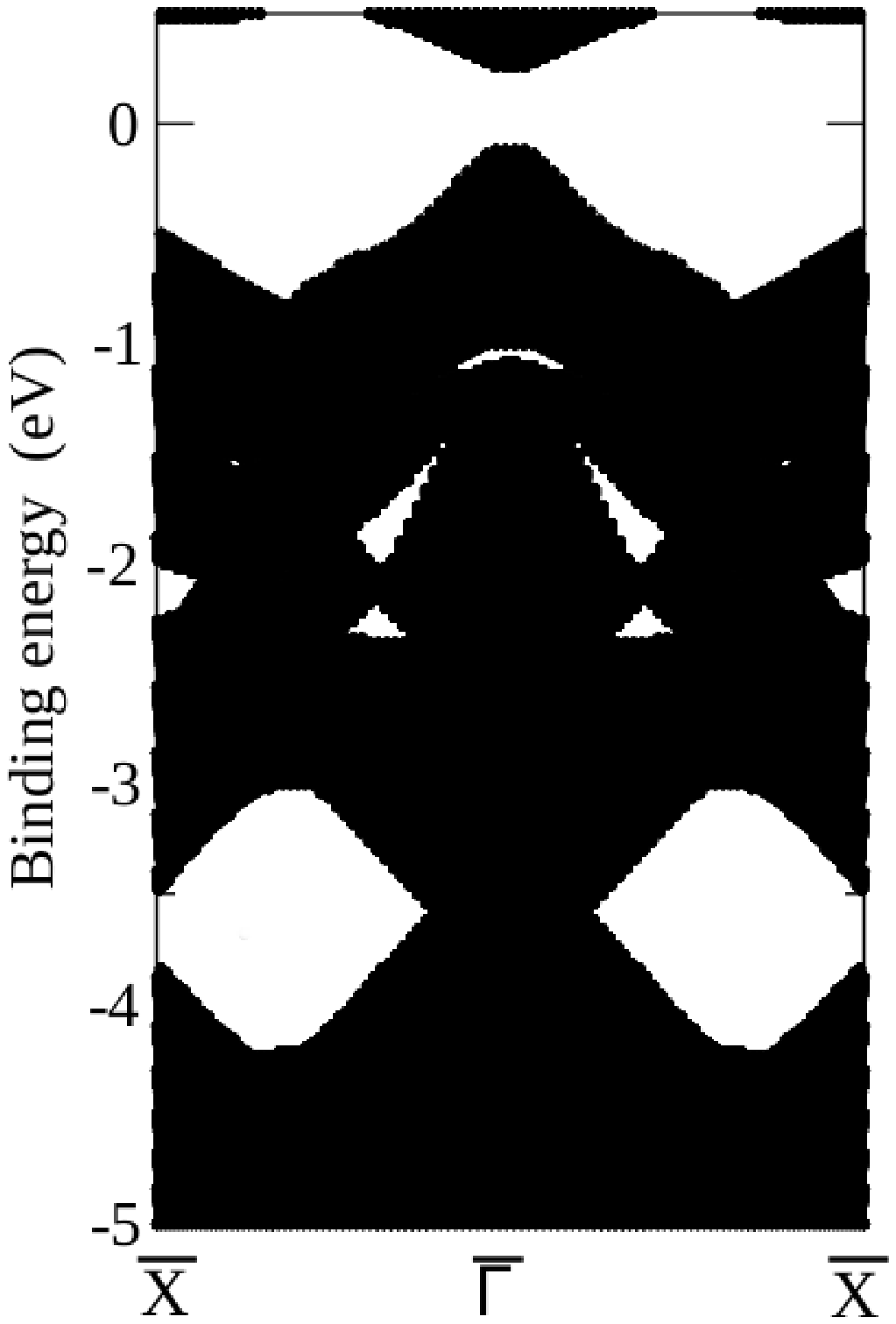}
\caption {Projected relativistic bulk band structures for majority spin (left panel) and minority spin character (right panel)
along $\overline \Gamma$-$\overline X$ direction. Dark colored regions represent the projection of bulk states.}
\label{figure3}
\end{figure}
Next we inspect the electronic structure of Co$_2$MnSi(100) along the two high symmetry directions 
$\overline \Gamma$-$\overline M$ and $\overline \Gamma$-$\overline X$ of the two-dimensional Brillouin zone. The bulk states of
Co$_2$MnSi projected along these two directions are shown in Figs.~2 and 3, where black color indicates bulk-like regions. The
left panels present the majority states and the right panels the corresponding minority states, with the total gaps visible
in the minority projected bulk-band structures around the Fermi level E$_{\rm F}$. Besides the total gaps appearing in the
bulk-related minority spin states no further gap structures are visible below E$_{\rm F}$. Only for binding energies higher
than about 1.5 eV symmetry-induced off-normal gaps exist. As a consequence one would not expect to find pronounced surface-related
features dispersing in this binding-energy regime. The situation is different for the unoccupied states. Relatively small gaps
appear in both minority spin-projected bulk-band structures just above the Fermi level for nonzero $\bf k_{||}$ values.
Furthermore, along $\overline \Gamma$-$\overline M$ larger off-normal gaps appear in the majority-spin projected band structure
near the $\overline M$ point, serving this way as an important precondition for the existence of surface resonances. The
problem with ground state electronic structure calculations is that surface-related features are often hidden in the continuum
of bulk states, because of their relatively small spectral weight. Their determination by use of photoemission calculations
is often more successful as surface resonances are typically enhanced in their spectral weight due to the excitation process.
This is a typical matrix element effect. Furthermore, the determinant criterion \cite{Bra04} allows for an additional check
on the surface contribution of a specific spectral distribution.

\begin{figure}[tp]
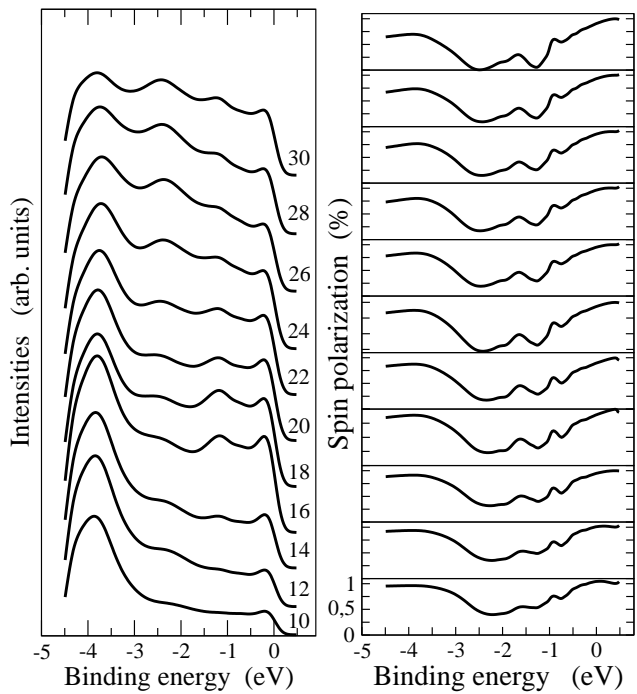

\includegraphics[width=0.23\textwidth,clip]{figure4a.eps}
\includegraphics[width=0.23\textwidth,clip]{figure4b.eps}
\caption {Left panel: Angle-integrated photoemission spectra calculated for different photon energies in the range from
h$\nu$ = 10 to 30 eV for p-polarized light. Right panel: In-plane component of the spin polarization vector
calculated along $\overline \Gamma$-$\overline M$.}
\label{figure4}
\end{figure}
In the following angle-integrated photocurrent calculations will be presented for the occupied states (AI-UPS), as well as for
the unoccupied states (AI-IPE), where IPE means inverse photoemission. Fig.~4 shows a series of spectra calculated for linear
p-polarized light as a function of the photon energy. The incidence angle of the incoming photon beam was chosen $\theta_p$=45$^o$
with respect to the surface normal. Just below the Fermi level a surface-related intensity distribution appears. This signal
has to be attributed to the majority surface resonance which is located about 0.4 eV above E$_{\rm F}$. Due to convolution with
a Fermi distribution function for room temperature and due to a Gauss folding with full width at half maximum of 0.2 eV the
tail of this spectral features seems to appear as a peak located at a finite binding energy \cite{Don89,Sto10}. At about 1.2 eV
binding energy a bulk-like signal is shown which could be attributed to Co and Mn majority d-states. The peak is visible over
the whole range of photon energies but with the highest intensity around 20 eV excitation energy. The dispersion is not very
pronounced with the tendency that the peak disperses to lower binding energies for higher photon energies. A third spectral
feature is found at about 2.5 eV binding energy, also with a less pronounced dispersion and relatively small variations in
the maximum intensity. These features also originate from majority Co and Mn-d states. The last spectral feature appears at about
4 eV binding energy and represents excitation from majority Co and Mn d-states, as well as from Si p-states. In the right panel
of Fig.~4 we present the corresponding spin polarization as a function of the excitation energy. First, one may observe that for
all photon energies the spin polarization reaches nearly 100$\%$ in the vicinity of the Fermi level. The spin polarization
decreases to about 40$\%$ for binding energies around 2 eV. For higher binding energies up to 4.5 eV the spin polarization
increases again and reaches high polarization values between 75$\%$ and 95$\%$, where the increase is more pronounced for
lower excitation energies. This result is in excellent agreement with corresponding experimental data available for photon
energies of h$\nu$=16.67 eV and  h$\nu$=21.2 eV, and will be discussed in more detail below. As a next step we want to
find out the reason for these unexpected high spin-polarization values at the Fermi level. To do so spin-resolved IPE spectra
have been calculated, again as a function of the photon energy. The result is presented in Fig.~5, where the left panel shows
the intensity distributions and the right panel the corresponding spin polarizations. Two dominant spectral features appear
for all excitation energies. The unoccupied majority surface resonance (SR), which is responsible for the spectral intensity
just below E$_{\rm F}$, and the series of image-potential states which are hidden in the broad peak at about 3.2 eV above E$_{\rm F}$.
The resonance is very intense and it is nearly 100$\%$ spin-polarized. This is clearly seen in the right panel of Fig.~5, where the
spin polarizations are shown. This way our analysis reveals that the pure bulk contribution of the experimental spin polarization,
which due to the limited experimental energy resolution was about 50$\%$ only, is increased to about 100$\%$ by the surface
resonance. 
\begin{figure}[bp]
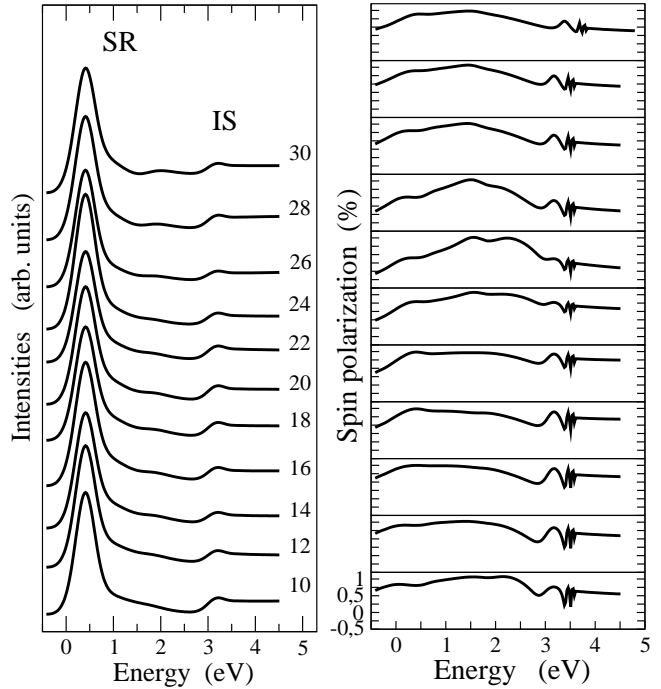

\includegraphics[width=0.23\textwidth,clip]{figure5a.eps}
\includegraphics[width=0.24\textwidth,clip]{figure5b.eps}
\caption {Left panel: Angle-integrated inverse photoemission spectra calculated for different photon energies in the range
from h$\nu$ = 10 to 30 eV for p-polarized light. Right panel: In-plane component of the spin polarization vector
calculated along $\overline \Gamma$-$\overline M$.}
\label{figure5}
\end{figure}

Even more, the amount of spectral weight and the high spin polarization value of this spectral feature are intimately
connected with the in-plane magnetization of the sample. In case where the magnetization is directed perpendicular to the
surface this resonance vanishes with a very low spectral weight into the bulk continuum, and instead one observes an occupied
minority surface state located just below E$_{\rm F}$. As a consequence the spin polarization at the Fermi level is reduced to values
significantly smaller than the pure bulk value. The origin for this peculiar behavior is found in the very different electronic
structures, which result for in-plane and out-of-plane magnetization directions of the sample surface. This is clearly seen,
if one inspects Fig.~1 again.

In Fig. 6 the spectroscopical calculations and the experimental spin-integrated UPS and HAXPES results are compared. Nearly
quantitative agreement for both, UV and hard X-ray photon energies, is obtained. Only the intensity distribution calculated
just below E$_{\rm F}$ at h$\nu$=16.67 eV is overestimated in comparison with the experimental data. The reason is found in the
energy-dependent cross section of the surface resonance which increases at lower photon energies because of the energy-dependent
multiple scattering between bulk and surface. As a consequence the wave function of the resonance is strongly energy-dependent
and so the corresponding matrix elements. Therefore, the theoretically overestimated spectral distribution may be ascribed to a
typical matrix-element effect. Besides this, and with regard to the small DOS just below the Fermi energy the agreement of the
calculations with the high UPS and HAXPES intensities in this energy range is remarkably good and it is mostly traced back to this
\begin{figure}[tp]
\hspace*{-0.8cm}
\includegraphics[width=0.57\textwidth,clip]{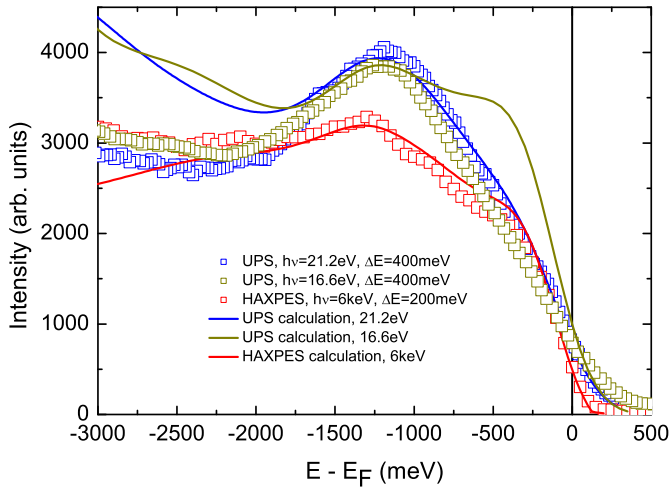}
\caption {Comparison of measured and calculated AI-UPS and angle-integrated HAXPES spectra. The HAXPES data have been
collected from ex situ capped Co$_2$MnSi thin layers with an energy resolution of $\Delta$E=200 meV for a photon energy of
h$\nu$=6 keV. The AI-UPS spectra were measured in situ on an uncapped Co$_2$MnSi layer with an energy resolution of $\Delta$E=400
meV for two different photon energies of h$\nu$=16.67 and 21.2 eV. The corresponding one-step photoemission calculations
are based on electronic structure calculations in the framework of the SPRKKR+DMFT method with U-parameters U$_{\rm Mn}$=3.0 eV and
U$_{\rm Co}$=1.5 eV for Mn and Co.}
\label{figure6}
\end{figure}
\begin{figure}[bp]
\hspace*{-0.8cm}
\includegraphics[width=0.57\textwidth,clip]{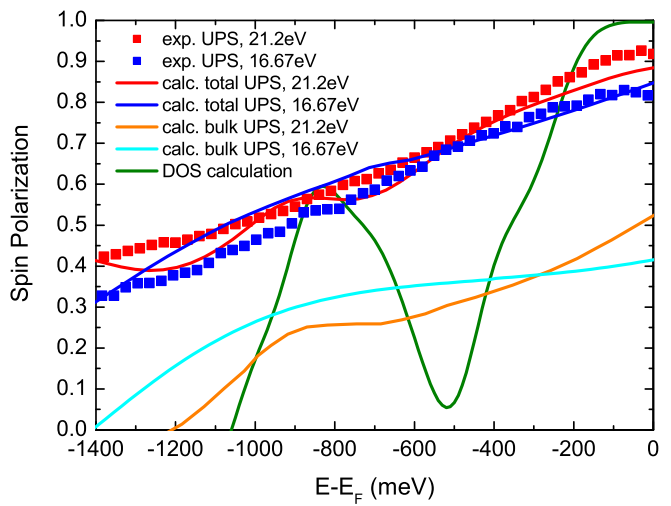}
\caption {Comparison of measured spin polarizations, taken in situ from an uncapped Co$_2$MnSi layer, with calculated ones.
Shown are the corresponding results for photon energies of h$\nu$=16.67 and 21.2 eV. In addition a bulk-like calculation
of the spin polarization is presented, where the surface contribution of the theoretically obtained photocurrent is suppressed.
The calculated DOS is shown in green (light) color, respectively. Data for h$\nu$=21.2 taken from \cite{Jou14}.}
\label{figure7}
\end{figure}
bulk-like surface resonance occuring in the majority-spin channel. The energetic position and dispersion behavior of this
spectroscopical feature has been discussed in detail above. Here it remains to remind to the fact that our self-consistent
electronic structure calculation leads to a half-metallic band structure with a total gap located around E$_{\rm F}$ in the minority
spin-projected states. But also from our experimental data strong evidence for half-metallicity is provided as we have estimated
the position of the lower band edge of the minority gap at about E-E$_{\rm F}$ = -0.5 eV, directly from the corresponding spectroscopical
data. As shown in Fig.~6, the inclusion of the complete surface-related photoexcitation in the UPS calculation results in nearly
perfect agreement with experiment. If the surface resonance were not present, half-metallic behavior would persist, but the
finite experimental resolution in photoemission would hinder the observation of a high spin polarization. As mentioned before,
the theoretical analysis reveals an experimental resolution limited spin polarization lower than 50$\%$ for a pure bulk-like
calculation. This provides further evidence for the calculated half-metallic band structure of Co$_2$MnSi.

In Fig.~7 the highest experimentally observed spin polarization is shown together with the calculated spin polarization for
two different photon energies of h$\nu$=16.67 and 21.2 eV, and with the corresponding theoretical DOS. The calculated 
photoemission asymmetries include all relevant broadening effects occuring in the measurements. In particular, the influence
of intrinsic life-time broadening generated by electronic correlations, broadening effects from impurity scattering and the
experimental resolution of about $\Delta E$=0.4 eV are considered. If one compares first the pure bulk-like theoretical spectrum
with the calculated DOS the correspondence between these two intensity distributions is obvious. The broadening effects in
combination with the absence of surface-like emission reduce the effective spin polarization tremendously, although
half-metallic behavior persists. Considering surface-related effects changes the situation dramatically. A true surface state
which typically disperses in a huge gap of a projected bulk-band structure shows up with a maximum spectral weight at the first
atomic layer. A well known example for such a surface feature is the Shockley state dispersing on the Cu(111)-surface \cite{Gra93}.
Therefore, the spectral weight compared to normal bulk states is small. Thus the combined effect of a very short inelastic
mean free path and an energy-dependent cross section reduces the spectral weight in the photoemission process significantly.
The situation is very different for Co$_2$MnSi because the majority surface resonance is embedded in the unoccupied bulk
continuum  with a strong coupling to the majority bulk states. This is because a layer-dependent analysis of the spectral 
weight showed that the resonance extends over the first six atomic layers of the semi-infinite bulk. This is similar to the
case of W(110), where we found a surface resonance revealing a considerable bulk contribution \cite{Bra14} as well. The spectral
weight of this surface resonance is much larger than that of a true surface state resulting in a significant contribution to
the total intensity even at hard X-ray energies.
\begin{figure}[tp]
\includegraphics[width=0.23\textwidth,clip]{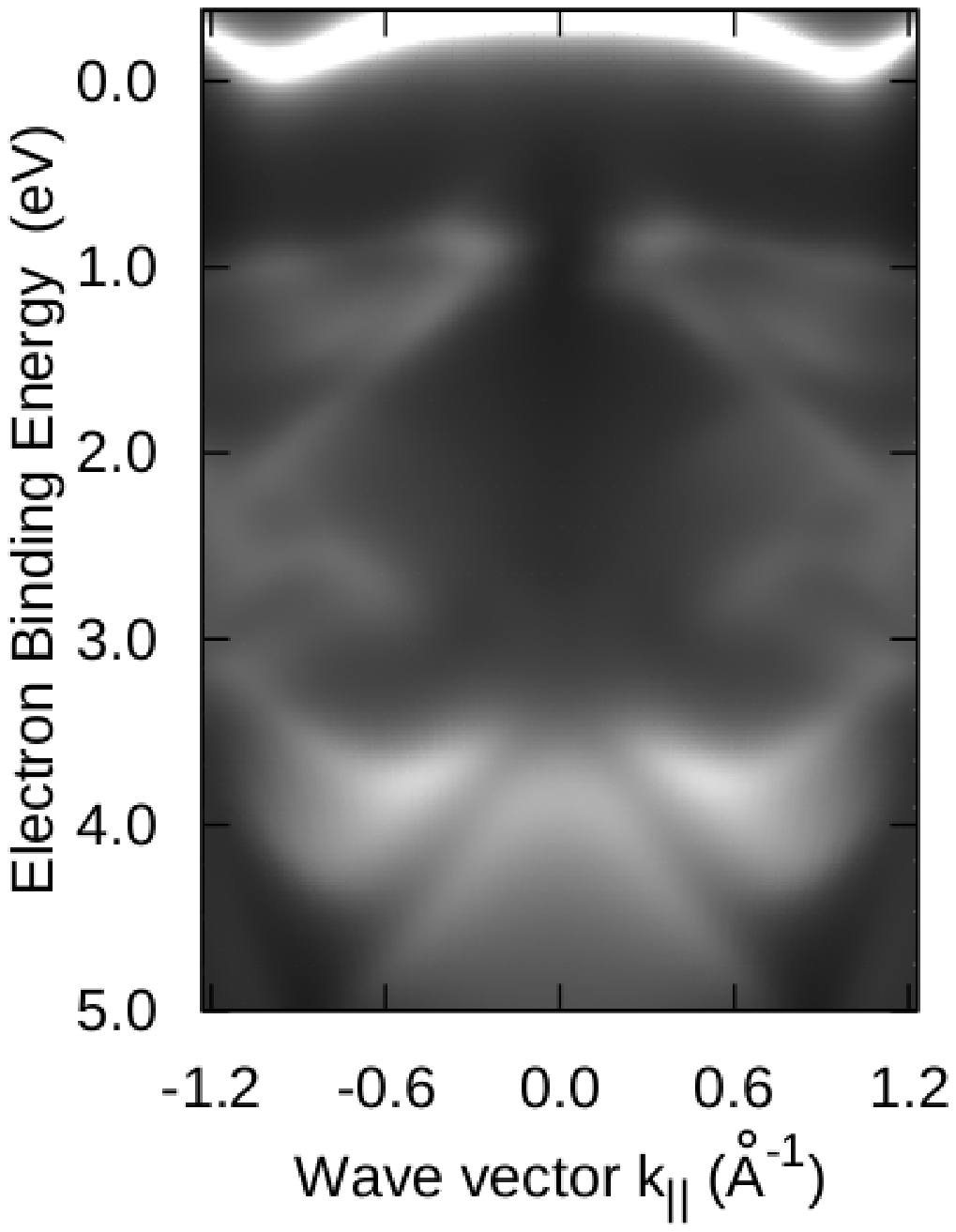}
\includegraphics[width=0.23\textwidth,clip]{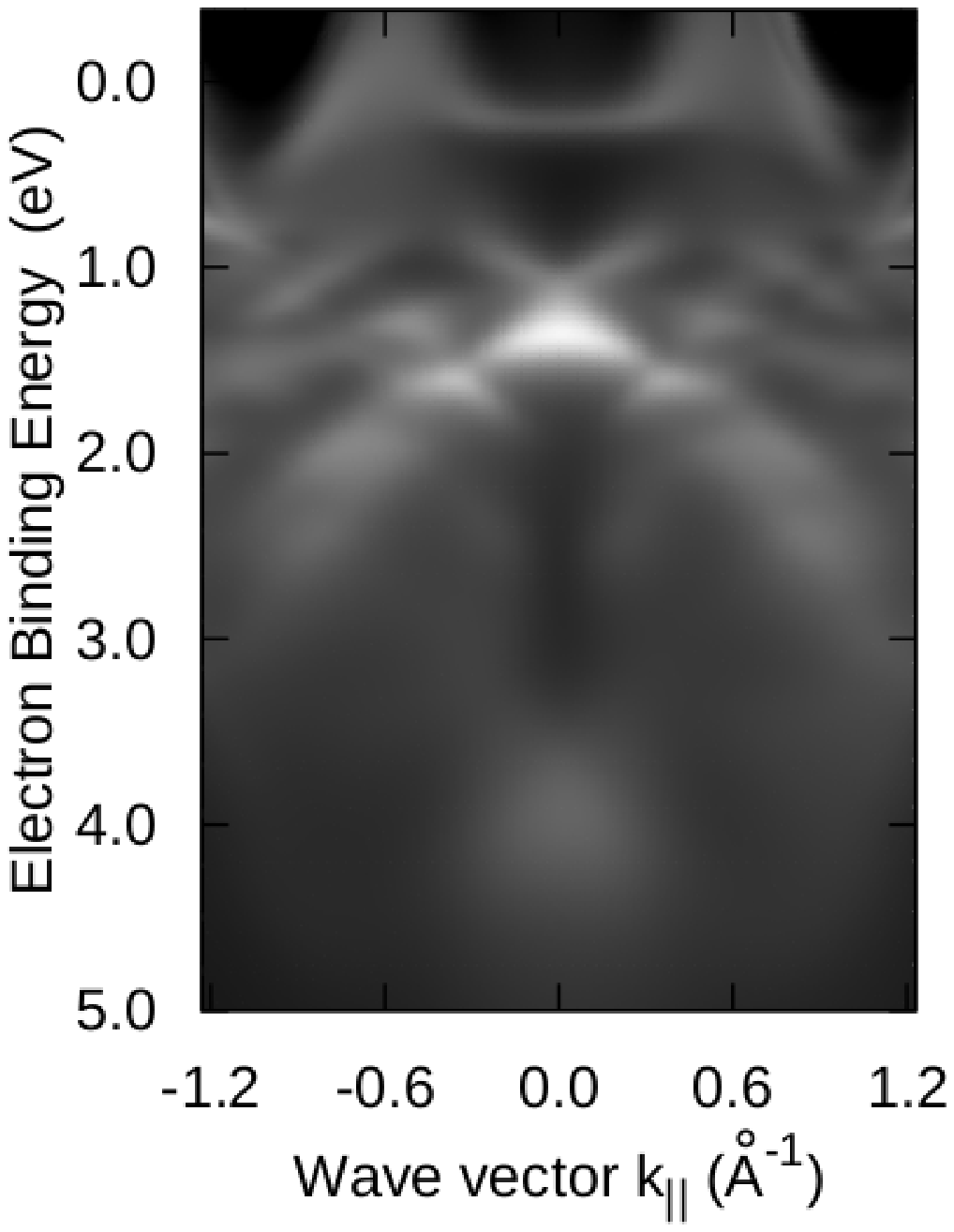}
\caption {Contour plots for majority (left panel) and minority (right panel) angle-resolved spectral densities are shown, which
have been calculated within the fully relativistic one-step model along the $\overline \Gamma$-$\overline M$ direction for a
photon energy of h$\nu$=21.2 eV. High spectral weight is indicated by light colors.}
\label{figure8}
\end{figure}

As a last point in our analysis we present spin-resolved ARPES spectra calculated as a function of binding energy and
$\bf k_{||}$-value along the $\overline \Gamma$-$\overline M$ direction of the surface Brillouin zone. The left panel of Fig.~8
shows the majority-spin intensity distribution in form of a contour plot. In the right panel the corresponding contour plot 
for minority-spin states is shown. High spectral weight is indicated by light colors. Not surprisingly, the highest spectral
weight belongs to the surface resonance. This is clearly observable from the majority contour plot. This feature slightly
disperses around $\overline \Gamma$ at about 0.3 eV above the Fermi level. Noticeable is here that this feature disperses
towards the Fermi level for higher $\bf k_{||}$-values and nearly touches E$_{\rm F}$ at $\bf k_{||} \approx$1.0 inverse
Angstr\"om. This result is not obtainable from angle-integrated photocurrent calculations. It supports even more our finding
that the resonance is able to enhance the experimental resolution limited spin polarization by almost a factor of two, although
this feature seems to be located quite far from the Fermi level if only calculated AI-UPS data are inspected. At about 0.5 eV
binding energy the bulk states show up to disperse as a function of E and $\bf k_{||}$. At lower binding energies in the region
between 0.5 and 3 eV mainly Co and Mn majority d-states are visible, where at higher binding energies around 4 eV a mixture of
Si p-states and Co and Mn d-states exists with stronger dispersion behavior. In agreement with our DOS calculations the spectral
features dispersing down to about 5 eV are more intense in their spectral signals than most of the states at lower binding energies.
This is due to the intense peak appearing in the DOS, which is ascribed to Si p-states, and is obviously a shortcoming of the
electronic structure calculation as discussed in detail in Ref.~\onlinecite{Cha09}.

In the right panel of Fig.~8 we present the minority states in form of an E versus $\bf k_{||}$ contour plot. Besides the
fact that the minority Mn and Co d-states are visible with their dispersion in $\bf k_{||}$, the most interesting observation
is the non-vanishing spectral density at the Fermi level. At $\bf k_{||}$-values of about $\pm$0.6 inverse Angstr\"om a nonzero
spectral signal is present. This is the fingerprint of a minority surface state which disperses into the total gap of the minority
spin-projected bulk-band structure. This feature is not visible in the AI-UPS spectra because of its low spectral weight.
The origin for this peculiar result is found in the in-plane magnetization of the sample. In fact this true surface state
appears with high spectral weight if the magnetization of the corresponding sample points perpendicular to the surface,
where the spectral weight of our surface resonance decreases strongly. The existence of this feature, even for a 100$\%$
in-plane magnetized sample surface is the most possible reason why the spin polarization is reduced by a few percent from
100$\%$ to about 93$\%$.

\section{Summary}
In conclusion, we were able to demonstrate half-metallic behavior for Co$_2$MnSi in combination with a nearly 100$\%$ spin
polarization at room temperature, directly measured and confirmed by our theoretical analysis. Our spectroscopical work 
has clearly demonstrated that the spin polarization very sensitively depends on the interplay of bulk- and surface-related
spectral features, where the magnetization direction of the sample surface plays a major role. In particular, we found
that the high spin polarization at the Fermi energy is related to a stable surface resonance in the majority-spin projected
states extending deep into the bulk of the sample. A description within the LSDA approach in combination with the DMFT method
results in a quantitative description of the electronic structure of Co$_2$MnSi. The use of a DMFT+LSDA electronic structure
calculation is important, whereas the application of the fully relativistic one-step model of photoemission in its spin-density
matrix formulation guarantees a quantitative analysis of the spectroscopical data. Our observations may serve as a useful 
information for future spintronic applications on the basis of Heusler alloys.

\begin{acknowledgments}
Financial support by the Deutsche Forschungsgemeinschaft through the SFB 925 (project B5), FOR1464 "ASPIMATT" (Project P 1.2-A), FOR 1346, 
Eb-154/23, Eb-154/26 and GSC 266 MAINZ as well as the EU (ERC-2007-StG 208162), is gratefully acknowledged. JM would like to acknowledge
the CENTEM project (reg. no. CZ.1.05/2.1.00/03.0088) and CENTEM PLUS (LO1402) cofunded by Ministry of Education, Youth and Sports of
Czech Republic. 
\end{acknowledgments}

\end{document}